\documentclass[aps,prl,twocolumn,showpacs,superscriptaddress,groupedaddress]{revtex4}  % for review and submission
\usepackage{graphicx}  % needed for figures
\usepackage{dcolumn}   % needed for some tables
\usepackage{bm}        % for math
\usepackage{amssymb}   % for math
\usepackage{natbib}
\begin{document}

% The following information is for internal review, please remove them for submission
\widetext
%\leftline{Version Round3 as of 7/10/2011}
%\leftline{Primary authors: Joe E. Physics}
%\leftline{To be submitted to (PRL, PRD-RC, PRD, PLB; choose one.)}
%\leftline{Comment to {\tt d0-run2eb-nnn@fnal.gov} by xxx, yyy}
%\centerline{\em INTERNAL DRAFT DOCUMENT -- NOT FOR PUBLIC DISTRIBUTION}

% the following line is for submission, including submission to the arXiv!!
%\hspace{5.2in} \mbox{Fermilab-Pub-04/xxx-E}

%"OLD TITLE"
%\title{Simple demonstration of the non-existence of a joint probability distribution for position and momentum variables}
%REVISED TITLE (NGJ):
\title{Breakdown of the classical description of a local system}
\author{Eran Kot}
\email[Corresponding author: ]{eran.kot@nbi.dk}
\affiliation{The Niels Bohr Institute, University of Copenhagen, DK--2100 Copenhagen, Denmark}
\author{Niels Gr{\o}nbech-Jensen}
\affiliation{The Niels Bohr International Academy, The Niels Bohr Institute, DK--2100 Copenhagen, Denmark}
\affiliation{Department of Applied Science, University of California, Davis, California 95616}
\author{Bo M.~Nielsen}
\affiliation{The Niels Bohr Institute, University of Copenhagen, DK--2100 Copenhagen, Denmark}
\author{Jonas S.~Neergaard-Nielsen}
\affiliation{The Niels Bohr Institute, University of Copenhagen, DK--2100 Copenhagen, Denmark}
\author{Eugene S.~Polzik}
\affiliation{The Niels Bohr Institute, University of Copenhagen, DK--2100 Copenhagen, Denmark}
\author{Anders S. S{\o}rensen}
\affiliation{The Niels Bohr Institute, University of Copenhagen, DK--2100 Copenhagen, Denmark}

\date{\today}

\begin{abstract}
%REVISED ABSTRACT (NGJ):
We provide a straightforward demonstration of a fundamental difference between classical and quantum mechanics for a single local system; namely the absence of a joint probability distribution of the position $x$ and momentum $p$. Elaborating on a recently reported criterion by Bednorz and Belzig [Phys. Rev. A {\bf 83},  52113] we derive a simple criterion that must be fulfilled for any joint probability distribution in classical physics. We demonstrate the violation of this criterion using homodyne measurement of a single photon state, thus proving a straightforward signature of the breakdown of a classical description of  the underlying state. Most importantly, the criterion used does not rely on quantum mechanics and can thus be used to demonstrate non-classicality of systems not immediately apparent to exhibit quantum behavior. The criterion is directly applicable any system described by the continuous canonical variables x and p, such as a mechanical or an electrical oscillator and a collective spin of a large ensemble.   

\end{abstract}

\maketitle

The conceptual differences between classical and quantum physics have intrigued and sometimes bewildered the physics community since the early days of quantum mechanics. This has led to a search for indisputable manifestations of the quantum world through observations of non-classical behavior in experiments. A field of particular curiosity is that of identifying the quantum to classical cross-over for ever larger systems, thereby eventually identifying non-classical effects in macroscopic systems. Recently this has led to the observation of, e.g., macroscopic entangled atomic ensembles \cite{RevModPhys.82.1041,Kimble2008}, interference of large molecules \cite{Arndt1999} and experiments pushing toward observing non-classical effects in mechanical oscillators \cite{Regal2008,Teufel2011,Chan2011}. 
In parallel to this fundamental interest, non-classicality is of central importance to quantum information processing, the essence of which is to advance computation beyond what is classically possible \cite{Nielsen2004}. However, in some instances quantum effects are claimed by
demonstrating consistency with an appropriate quantum model. Yet any rigorous demonstration of genuine quantum behavior must exclude the possibility of classical explanations. The importance of this is exemplified in Ref.~\cite{GronbechJensen2010}, where a pair of coupled classical oscillators is shown to exhibit signatures easily mistaken for those of entanglement expected from a quantum model. Thus, a definite conclusion on the quantum nature of a system can only result from the breakdown of the classical description and not from verified agreement with quantum mechanics. This approach is most rigorously demonstrated by the Bell-inequalities,
  where the underlying model of the system is stripped of any physics and is reduced to the very basic assumptions of locality and realism, resulting in an indisputable non-classicality criterion. The Bell-inequalities cannot however, by their very nature, be investigated by data obtained from a single system.

In this letter we provide a conceptually simple demonstration of one of the key discrepancies
between classical and quantum mechanics, valid for systems of a single degree of freedom: classical systems can always be described by a joint probability distribution for $x$ and $p$, the two canonically conjugated coordinates of a system, whereas such a description does not apply in quantum mechanics due to the Heisenberg uncertainty principle. This discrepancy is most evident when the phase space description of the state of a system is examined. Classically, the phase space distribution $W(x_i,p_i)$ is the joint probability of finding
the system in an infinitesimal area around $x=x_i, p=p_i$%
, and hence it obeys all the requirements of a probability distribution including being a non-negative function. As mentioned, in the case of a quantum phase space formulation, introduced by Wigner \cite{Mandel1995}, the Heisenberg uncertainty renders this definition meaningless, as a joint probability distribution for $x$ and $p$ does not exist.
The phase space distribution is only defined through the single coordinate (marginal) distributions, projected from the distribution function \cite{Nha_PRA.78.012103} and this relaxation of constraints allows for negative values of the function in areas smaller than $\hbar$. This negativity is not directly observable due to the vacuum fluctuations preventing simultaneous measurement of $x$ and $p$. However, one can still infer the phase space distribution from measurements of only a single observable at a time and detect such negativities, thereby illuminating the failure of classical theory.

The usage of these negativities as markers of non-classicality has been discussed and demonstrated in several quantum optics systems (see, e.g., \cite{Leibfried1996,Lvovsky2001,Zavatta2004,Ourjoumtsev2006,Mari2011}), using tomographic techniques.
Often such methods search for the quantum state most compatible with the experimental data
using statistical inference  or variational techniques\cite{Hardil1997,Furusawa2011} and thus inherently rely on quantum mechanics. These methods are therefore not applicable for demonstrating the absence of a classical description. Alternatively, given measurements of all the coordinate distributions, the underlying state can be uniquely determined, and the phase space distribution fully calculated using the inverse Radon transformation \cite{Welsch199963} without relying on quantum mechanics.  
Though such methods have been used in quantum optics for demonstrating various states, the mathematical transformation involved is highly complicated. Furthermore the numerical stability of the inverse transformation is problematic, leading to numerical uncertainty at high frequencies, and sometimes results in unphysical states \cite{RevModPhys.81.299}.
These limitations are a drawback for using tomographic techniques for validating the breakdown of a classical description, and the application of these methods is usually cumbersome.

Our simple, unambiguous demonstration of the absence of a classical probability distribution is based on recent theoretical work by Bednorz and Belzig \cite{Belzig_PRA.83.052113} that verifies the negativity of the Wigner function based on moments. 
As discussed in detail below, their results lead to a hierarchy of inequalities, such that violation of any one inequality indicates negativity of the Wigner function. Full tomographic reconstruction with the associated numerical complexities is thereby avoided. We extend this approach such that it can be applied to quadrature measurement of a single photon state, and use the experimental data from the heralded single photon generation to directly disprove the existence of a joint probability of the position and momentum for this system.

We start by re-iterating the key results of Bednorz and Belzig, through a reformulation that relies only on classical mechanics. The phase space of a system with a single degree of freedom is fully characterized by a two-dimensional phase space distribution $W(x,p)$. That is, given the phase space distribution, the ensemble averaged result of any measurable quantity $A$ can be obtained by
\begin{equation}
\langle A\rangle=\int\!dx dp\, W(x,p)\,A(x,p) \, ,
\label{eq:AvgDef}\end{equation}
where $A(x,p)$ is the decomposition of the quantity $A$ in terms of the generalized coordinate $x$ and its canonically conjugated momentum $p$.

To disprove the existence of a classical probability distribution we examine the ensemble average of a non-negative test function $\mathfrak{F}(x,p)$ over a classically explainable system, which must have a proper distribution function that results in the ensemble average of $\mathfrak{F}$ be non-negative:
\begin{equation}
\langle \mathfrak{F}\rangle=\int\!dx dp\, W(x,p)\,\mathfrak{F}(x,p)\geq 0.
\label{eq:require}
\end{equation}
Violating this condition is a direct proof of the absence of a joint probability distribution. The condition can, however, be violated in quantum mechanics, where $W(x,p)$ is the Wigner function that can contain negative values. The objective therefore is to optimize a test function such that it will be dominant at the possible negative areas of the distribution function.
For a rotationally invariant phase space
both the phase space distribution and the test function can be described solely by the phase space radius $r$, defined by $r^2 = x^2+p^2$. For reasons to become clear later, we choose a specific form for the test function $\mathfrak{F}$, writing it as a square of an $N$th order, even polynomial $M$ with real coefficients $\{C_{i}\}_N$;
\begin{equation}
\langle \mathfrak{F}\rangle = \langle M^2\rangle=\left\langle\left(1+\sum^{N/2}_{n=1}C_{2n}r^{2n}\right)^2\right\rangle
\label{eq:functional}
.\end{equation}
Minimizing the above expression for a given order $N$ is done by straight-forward linear optimization of the coefficients $\{C_{i}\}_N$:
\begin{equation}
\sum_{l=1}^{N/2}\langle r^{2(l+j)} \rangle C_{2l} = - \langle r^{2j} \rangle
\label{eq:optimazation}
,\end{equation}
for all $j=1,2,...,N/2$. Notice that the linearity of the problem ensures that the obtained minimum of $\langle\mathfrak{F}\rangle$ is global and therefore the most optimal indicator of a possible violation of Eq.~(\ref{eq:require}) for a given polynomial order $N$.
It is important to emphasize that this is only a sufficient criterion for non-classicality, and an optimized positive average for a chosen $N$ does not ensure a classical probability distribution, since the negativity may only be exhibited by the inclusion of higher order terms in $M$. However, it is clear that increasing the polynomial order $N$ cannot increase the minimized value of $\langle\mathfrak{F}\rangle$, and we conjecture that the limit of $N\rightarrow\infty$ will exhibit any negativity of the Wigner function, as the polynomial can represent an arbitrarily (analytical) sharp peaked function $\mathfrak{F}$ focused at the negativity. Assuming the existence of all moments (e.g., due to an exponentially decaying tail of the phase space distribution at large $r$), this then becomes a necessary criterion for the negativity of the distribution function. We also note here, that similar polynomial expansion has been discussed \cite{Shchukin2004,Korbicz2005}
in the context of the P-function distribution. 
The P-function is, however, only defined within the framework of quantum mechanics, and hence cannot be used to prove the absence of a classical description.

We assume that, as is the case for many systems, the system in question can only be experimentally accessed by measuring one of the canonically conjugated variables (e.g., $x$ or $p$) at a time. Since we are restricted to single coordinate measurement at a time, neither the intensity nor the phase space distribution function is directly accessible. For this method to be applicable to such experimental data, the functional $\langle \mathfrak{F}\rangle$ must be expressed in terms of the moments of the projected coordinates  $\langle Q_{\alpha}^n\rangle$, where 
\begin{equation}
Q_\alpha=\cos\!\alpha\,x+\sin\!\alpha\,p,\; P_\alpha= \cos\!\alpha\,x-\sin\!\alpha\,p \, 
\label{eq:quadratures}
\end{equation} 
is a measureable rotated coordinate. 
To do this we use the identity
\begin{eqnarray}
\left(x^2\!+\!p^2\right)^{N} =
A_{2N}\!\sum_{m=1}^{2N}\!\left(\cos\!\left(\frac{m\pi}{2N}\right)x\!+\!\sin\!\left(\frac{m\pi}{2N}\right)p\right)^{2N},
\label{eq:mathIdentity}
\end{eqnarray}
where
\begin{eqnarray}
A_{2N} = \left(
\begin{array}{c}
2N\\
N
\end{array}\right)^{-1}\frac{2^{2N}}{2N}.
\end{eqnarray}
This is where quantum and classical approaches diverge. While classically Eq.~(\ref{eq:mathIdentity}) represents a measurable physical quantity, it is missing the key vacuum uncertainty, allowing for the breakdown of the classical description.

It is interesting to note the implication of identity (\ref{eq:mathIdentity}). For the $2m$th moment of the radial distribution to be known, we need $2m$ 'cuts' in phase space; i.e., different coordinate measurements at equally distributed angles. Regardless of any assumption about the underlying state, the average of  Eq.~(\ref{eq:mathIdentity}) directly gives
\begin{eqnarray}
\langle r^{2N}\rangle =
A_{2N}\sum_{m=1}^{2N}\left\langle \left(Q_{m\pi/2N}\right)^{2N}\right\rangle.
\label{eq:moments}
\end{eqnarray}
In the special case of a symmetric distribution function these moments are all identical, and Eq.~(\ref{eq:moments}) reduces to
\begin{eqnarray}
\langle r^{2N}\rangle=\left(
\begin{array}{c}
2N\\
N
\end{array}\right)^{-1}2^{2N}\langle x^{2N}\rangle.
\end{eqnarray}
The radial moments can thus be indirectly calculated from the quadrature measurements. Substituting these radial moments into Eq.~(\ref{eq:require}) using the functional form of $\mathfrak{F}(x,p)$ given by Eq.~(\ref{eq:functional}), we get, for a given set of measured moments $\{\langle x^{2k}\rangle\}_k$, a necessary condition for classicality of the underlying state. If Eq.~(\ref{eq:require}) is violated by the solutions of Eq.~(\ref{eq:optimazation}), the underlying state cannot be explained by a proper phase space probability distribution, and one cannot assign a joint probability distribution to $x$ and $p$. 

\begin{figure}
\includegraphics[scale=0.25]{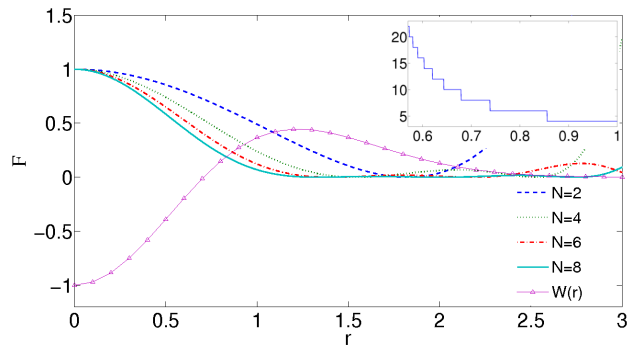}
\caption{\label{fig:epsart2} Profiles of the test function $\mathfrak{F}$ minimizing the expectation value $\langle\mathfrak{F}\rangle$ for the first excited state of a quantum harmonic oscillator, as a function of the phase space radius, for different orders $N$ (see text) plotted against the profile of the corresponding Wigner function. As the order of the polynomial increases, the function becomes centered around the negativity, decreasing elsewhere. In this case, negative expectation values are obtained starting at $N=4$. Inset shows the polynomial order required to observe negative expectation values, as a function of the single-photon fractional content in a mixture with vacuum. As the fraction of vacuum is increased, the state approaches a classically describable state and higher moments are needed to observe the negativity.}
\end{figure}

To demonstrate the absence of a joint probability distribution we are going to consider the phase space description of a single photon state. In phase space this can be described by the first excited state of a harmonic oscillator, which is rotationally invariant and contain negative parts in the Wigner functions. Fig.~\ref{fig:epsart2} shows the optimal functional forms obtained for the this state for low polynomial orders. As higher order terms are included, the optimized test function is increasingly probing the negative part of $W$, yielding a negative expectation value.
We note that negative expectation values appear only from the fourth order onward. This is because the peak of the test function at the position of the negativity must be narrower than Heisenberg's uncertainty in order not to smear the negativity; this is in full agreement with Ref.~\cite{Belzig_PRA.83.052113}.

The experimental demonstration is achieved with single photons generated by an heralded cavity-enhanced non-degenerate parametric down-conversion. The equivalence between a single mode electromagnetic field and an harmonic oscillator allows us to describe the EM field by a phase space of a single degree of freedom. The down-conversion process produces two photons, and as one is detected as a trigger, the result is a single photon state where the losses introduce a statistically mixed component of vacuum. The projection measurements (quadratures) are obtained by measuring the statistics of the noise, using an optical homodyne detection scheme. In this scheme, the weak investigated optical field is overlapped with a strong laser pulse on a beam splitter, and the interference of the two fields is detected and subtracted. The phase of the strong laser field determines the angle $\alpha$ (Eq.~(\ref{eq:quadratures})) of the measured coordinate. Measurements were taken without fixing the phase of the local oscillator, thus smearing the resulting distribution. This enables us to treat the results as rotationally invariant even if non-invariant features existed prior to smearing. Such measurements will generate a rotationally invariant reconstructed state for any underlying state, but this does not necessarily average out negativities in the Wigner function (see, e.g., \cite{Leonhardt2005}). For details of the experimental setup and the charactarization of the resulting single photon see Ref.~\cite{Neergaard-Nielsen:07}.

\begin{figure}
\includegraphics[scale=0.25]{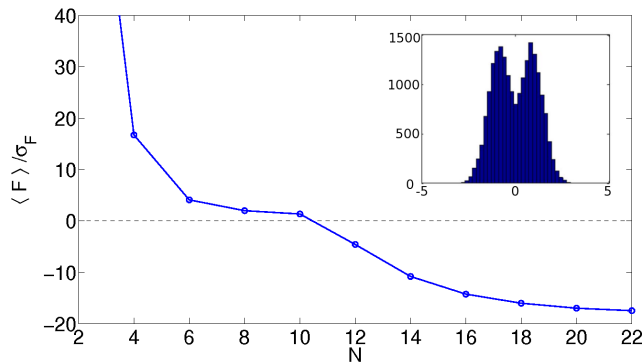}
\caption{\label{fig:epsart} Expectation value for the square of a polynomial relative to its standard deviation, as a function of the polynomial's order for the experimental data. Negativity by almost $20$ standard deviations disproves the existence of a joint probability distribution for $x$ and $p$. The inset shows a histogram of the raw measured quadrature data (arbitrary units).}
\end{figure}

The data set contained 180,000 measured quadratures. We have here revised the optimization of the functional to also account for statistical uncertainties inherent to a limited data set. This is done by optimizing
\begin{equation}
G=\frac{\langle \mathfrak{F}\rangle}{\langle \sigma_{\mathfrak{F}}\rangle},
\end{equation}
where $\sigma_{\mathfrak{F}}=\sqrt{\langle\mathfrak{F}^2\rangle-\langle\mathfrak{F}\rangle^2}$ is the standard deviation of $\mathfrak{F}$. The results are shown in Fig.~\ref{fig:epsart}. The fact that the expectation value for our test function is negative with certainty of almost twenty standard deviations, clearly demonstrates that the measured state in this experiment cannot be explained by classical theory, unambiguously negating the possibility of existence of a joint probability distribution for $x$ and $p$. The appearance of negative values from the twelfth order polynomials and higher indicate the quantum mechanical description of this state in terms of a Wigner function includes negative valued areas. We note that the minimized function from Eq.~(\ref{eq:functional}) is monotonically decreasing for increasing order $N$, and the onset of negativity at a certain order therefore means that all higher orders will also be negative. This suggests a sequential authentication procedure for an unknown state.
As mentioned above, for a pure single photon state, negative expectation values are observable from the $4$-th order polynomial onwards. The twelfth order polynomial required here is due to the vacuum component of the field, requiring higher orders of the polynomial as shown in the inset of Fig.~\ref{fig:epsart}, and is in agreement with the results obtained in Ref.~\cite{Neergaard-Nielsen:07} reporting 62\% fraction of single photon in the resulting mixed state.

In conclusion, we have experimentally demonstrated the non-existence of a joint probability distribution of two canonical variables. This is done by violation of an inequality derived without the assumptions of quantum mechanics, thus allowing for it as proof of the absence of a classical description in systems not immediately evident to display quantum behavior. The procedure used here can thus provide a simple, practical tool for demonstrating the non-classicality of a state based on quadrature measurements, where the existence of a classical joint distribution of two conjugated variables can be negated. In this way, this procedure is closely linked to other criteria \cite{Legget1985,Klyachko2008,Lapkiewicz2011} demonstrating contextuality of measurements, and thus disproving the classical local hidden variable view. Unlike Ref.~\cite{Legget1985,Klyachko2008,Lapkiewicz2011}, which are applicable to discrete variables, the method demonstrated here applies for continuous variables such as position and momentum, collective spin operators \cite{Fernholz2008} and quadrature phase operators. This makes it useful to systems containing many particles, where criteria based on counting particles are not easily implemented and interpreted. This method complements the full tomographic reconstruction techniques in that it is simpler and avoids numerical complexities of inverse transformations. These kinds of conceptual proofs, when extended to different detection schemes, can shed more light on the quantum to classical correspondence, especially where the control of claimed macroscopic quantum states is in question.

\section{Acknowledgment}
The authors would like to thank W. Vogel for his valuable comments and discussions. This work was support by the Villum Kann Rasmussen Foundation (Denmark), the EU projects Q-ESSENCE and MALICIA, and Danish National Research Foundation.
NGJ is grateful for generous support from Danmarks Nationalbank (Denmark).
% \bibliographystyle{prsty1}
% \bibliography{NonClassicNew}

\end{document}